\documentstyle[preprint,aps,epsf,epsfig,psfig,amssymb]{revtex}
\tighten
\begin{document}
\title{Bound States and Scattering Processes
       in the $^4$He$_3$ Atomic System}
\author{ A. K. Motovilov\thanks{On leave of absence from
	 the Laboratory of Theoretical Physics,
    Joint Institute  for Nuclear Research, Dubna, 141980, Russia},
	S. A. Sofianos}
\address{Physics Department, University of South Africa
            P.O.Box 392, Pretoria 0001, South Africa}
\author{E. A. Kolganova}
\address{ Laboratory of Computing Techniques and Automation,
Joint Institute  for Nuclear Research, Dubna, 141980, Russia}
\maketitle
\begin{abstract}
We present a mathematically rigorous method for solving three-atomic
bound state and scattering problems. The method is well suited
for applications in systems where the inter-atomic interaction
is of a hard-core nature. It has been employed to obtain
the ground- and excited-state energies for the Helium trimer and
to calculate, for the first time, the scattering phase shifts
and wave-functions for the He atom--He dimer at ultra-low energies.
\bigskip

\noindent LANL E-print physics/9709037;

\noindent Published in Chem. Phys. Lett. {\bf 275} (1997), 168--172.
\end{abstract}
\vspace*{1cm}

The $^4$He triatomic system is of interest in various
areas of physical chemistry and molecular physics.  The study of
the Helium dimer and trimer properties is the first step towards
the understanding of the Helium liquid drops, superfluidity in $^4$He
films, finite pores~\cite{RamaKrishna} {\it etc.}

Various theoretical and experimental works have been devoted in
the past to study the ground state properties of the $^4$He
clusters. From the theoretical works we mention here those using
Variational and Monte Carlo type methods~\cite{V1,V2,V3,V4,V5},
the Faddeev equations~\cite{Nakai,Gloeckle,CGM}, and the
hyperspherical approach~\cite{Levinger,Sof,EsryLinGreene}. From
the experimental works we recall those of
Refs.~\cite{DimerExp,DimerExp1,Science,Toennis2} where the
Helium dimer and trimer clusters were investigated.

Despite the efforts made to solve the He-trimer problem various
questions such as the existence of Efimov states and the study
of scattering processes still have not been satisfactorily
addressed. In particular  for scattering processes there are no
works which we are aware of apart from a zero-energy calculation
of Ref.~\cite{Nakai} and a recent study~\cite{Fed96} concerning
recombination rates. There are various reasons for this, the main one
being that the three-body calculations involved are extremely
difficult to perform due to the practically hard-core of the
interatomic interaction which gives rise to strong numerical
inaccuracies that make calculations cumbersome and unstable.

In this work we employed a mathematically rigorous method based
on a hard-core version~\cite{MerMot,MMYa} of the
boundary-condition model to calculate the binding energies and the
ultra-low energy scattering phase shifts below as well as above
the breakup threshold. Such an approach  takes into account,
from the beginning, the hard-core nature of the \mbox{He--He}
interatomic interaction. We show that this method is highly successful
and suitable for solving three-body bound state and scattering problems
in configuration space when the two-body interactions have a
hard-core.

In the present investigation we consider that the $^4$He$_3$ molecule
has a total angular momentum $L=0$. In this case one has to solve the
following, two-dimensional, integro-differential Faddeev equations~\cite{MF}
\begin{equation}
\label{FadPart}
   \left[-\displaystyle\frac{\partial^2}{\partial x^2}
            -\displaystyle\frac{\partial^2}{\partial y^2}
            +l(l+1)\left(\displaystyle\frac{1}{x^2}
            +\displaystyle\frac{1}{y^2}\right)
    -E\right]\Phi_l(x,y)=\left\{
            \begin{array}{cl} -V(x)\Psi_l(x,y), & x>c \\
                    0,                  & x<c\,.
\end{array}\right.
\end{equation}
Here, $x,y$ stand for the standard Jacobi variables
and $c$, for the core range.  The angular momentum  $l$ corresponds
to a dimer subsystem and a complementary atom; for the
$S$-state three-boson system $l$ is even, $l=0,2,4,\ldots\,.$
$V(x)$ is the He-He central potential acting
outside the core domain. The partial wave function $\Psi_l(x,y)$ is
related to the Faddeev components  $\Phi_l(x,y)$ by
\begin{equation}
\label{FTconn}
         \Psi_l(x,y)=\Phi_l(x,y) + \sum_{l'}\int_{-1}^{+1}
         d\eta\,h_{l l'}(x,y,\eta)\,\Phi_{l'}(x',y')
\end{equation}
where
$$
          x'=\sqrt{\displaystyle\frac{1}{4}\,x^2+\displaystyle
    \frac{3}{4}\,y^2-\displaystyle\frac{\sqrt{3}}{2}\,xy\eta}\,,
\qquad
         y'=\sqrt{\displaystyle\frac{3}{4}\,x^2+\displaystyle
   \frac{1}{4}\,y^2+ \displaystyle\frac{\sqrt{3}}{2}\,xy\eta}\,,
$$
and  $1 \leq{\eta}\leq 1$. The explicit form of the function
$h_{ll'}$ can be found in Refs.~\cite{MF,MGL}.

The functions $\Phi_{l}(x,y)$ satisfy the boundary conditions
\begin{equation}
\label{BCStandard}
      \Phi_{l}(x,y)\left.\right|_{x=0}
      =\Phi_{l}(x,y)\left.\right|_{y=0}=0\,.
\end{equation}
In the hard-core model these functions satisfy also the condition
\begin{equation}
\label{BCCorePart}
       \Phi_{l}(c,y) + \sum_{l'}\int_{-1}^{+1}
       du\,h_{l l'}(c,y,\eta)\,\Phi_{l'}(x',y')=0\,
\end{equation}
requiring the wave function  $\Psi_{l}(x,y)$ to be zero on the
core boundary $x=c$. In fact, one can show that, in general, the
condition (\ref{BCCorePart}) causes the wave functions~(\ref{FTconn})
to vanish inside the core domains as well. Moreover, for
the helium trimer bound-state problem the functions $\Phi_{l}(x,y)$
satisfy as $\rho\rightarrow\infty$ and/or
$y\rightarrow\infty$ the asymptotic condition
\begin{equation}
\label{HeBS}
    \begin{array}{rcl}
 \Phi_l(x,y) & = & \delta_{l0}\psi_d(x)\exp({\rm i}
 \sqrt{E-\epsilon_d}\,y)
 \left[{\rm a}_0+o\left(y^{-1/2}\right)\right] \\
        & + &
     \displaystyle\frac{\exp({\rm i}\sqrt{E}\rho)}{\sqrt{\rho}}
     \left[A_l(\theta)+o\left(\rho^{-1/2}\right)\right]
\end{array}
\end{equation}
where $\epsilon_d$ is the dimer energy and $\psi_d(x)$, the
dimer wave function.  The $\rho$, $\rho=\sqrt{x^2+y^2}$\,,
and $\theta$, $\theta=\arctan\displaystyle\frac{y}{x}$\,, are
the hyperradius and hyperangle for the trimer. The coefficients ${\rm
a}_0$ and $A_l(\theta)$ describe contributions into $\Phi_l$
from the $(2+1)$ and $(1+1+1)$ channels respectively.
It should be noted that both the $E-\epsilon_d$ and $E$
in~(\ref{HeBS}) for a bound state are strictly negative. This
implies that for any  $\theta$ the function $\Phi_l$ is
exponentially decreasing in $\rho$ as $\rho\to\infty$ .

The asymptotic boundary condition of the partial Faddeev
components for the $(2+1\rightarrow 2+1\,;\,1+1+1)$ scattering
wave function as $\rho\rightarrow\infty$ and/or
$y\rightarrow\infty$ reads
\begin{equation}
\label{AsBCPartS}
    \begin{array}{rcl}
      \Phi_l(x,y;p) & = &
      \delta_{l0}\psi_d(x)\left\{\sin(py) + \exp({\rm i}py)
      \left[{\rm a}_0(p)+o\left(y^{-1/2}\right)\right]\right\} \\
      & + &
  \displaystyle\frac{\exp({\rm i}\sqrt{E}\rho)}{\sqrt{\rho}}
                \left[A_l(\theta)+o\left(\rho^{-1/2}\right)\right]
    \end{array}
\end{equation}
where $p$ is the relative momentum conjugate to the variable $y$,
 E is the scattering energy given by $E=\epsilon_d+p^2$, and
 ${\rm a}_0$ is the elastic scattering amplitude. The $S$-state
elastic scattering phase shifts $\delta_0(p)$ are then given by
$$
    \delta_0(p)=\frac{1}{2}\,{\rm Im}\,\ln {\rm S}_0(p)
$$
where ${\rm S}_0(p)=1+2i{\rm a}_0(p)$ is the
$(2+1{\rightarrow}2+1)$ partial component of the scattering
matrix.  The functions $A_l(\theta)$ provide us, at $E>0$, the
corresponding partial Faddeev breakup amplitudes.

We employed the Faddeev equations~(\ref{FadPart}), the hard-core 
condition  (\ref{BCCorePart}), and the  asymptotic expressions 
(\ref{HeBS}, \ref{AsBCPartS}), to calculate the binding energies 
of the Helium trimer and the ultra-low energy phase shifts of 
the Helium atom scattered by  the Helium diatomic molecule. In 
our calculations we used $\hbar^2/m=12.12$\,K\,\AA$^2$.  Our 
finite-difference algorithm was closed in essential to that 
described in~\cite{MF,MGL}.  As a $^4$He--$^4$He interaction we 
employed the HFDHE2~\cite{Aziz79} and HFD-B~\cite{Aziz87} 
potentials of Aziz and co-workers which we found that they 
sustain a dimer bound state at $-0.8301$\,mK and $-1.6854$\,mK 
respectively.  The corresponding  $^4$He atom--$^4$He atom 
scattering length was found to be 124.7\,{\AA} for the HFDHE2 
and 88.6\,{\AA} for the HFD-B potential.

The results of the Helium trimer ground-state energy
calculations are presented in Table~I. Although the two
potentials used  differ only slightly, they produce  important
differences in the ground-state  energy. This is in agreement
with the finding of Ref.~\cite{Sof} but in disagreement with the
statement made in Ref.~\cite{V5}.  It should be further noted
that most of the contribution to the binding energy stems from
the  $l=0$ and $l=2$ partial component the  latter being more
than 35\,\%. The contribution from the $l=4 $ channel was shown
in~\cite{CGM} to be of the order of a few per cent. We have
found that the Helium trimer can form an excited state with both
the HFDHE2 and HFD-B potentials in agreement with the findings
of Refs.~\cite{Nakai,Gloeckle,EsryLinGreene}.  Note that in
Refs.  \cite{Gloeckle,EsryLinGreene} this state was interpreted
as an Efimov one~\cite{VEfimov}. Our excited state results are
given in Table~II.

The phase shift results for a Helium atom scattered by a Helium
dimer are plotted in Fig.~1.  We considered incident energies
below as well as above the breakup threshold, i.e., for the
($2+1{\longrightarrow}2+1$) and the
($2+1{\longrightarrow}1+1+1$) processes. It is seen that,
similarly to the bound state results, the inclusion of the $l=2$
partial wave is essential to describe the scattering correctly.
The relevant partial wave functions $\Psi_l(x,y;p)$, $l=0,2,$
calculated at $E=4.1$\,mK with the inclusion of both channels
$l=0$ and $l=2$ are plotted in Figs.~2--5.

Further to the bound and scattering calculations we endeavour to
estimate the scattering length
$$
 \ell_{\rm sc}=-\frac{\sqrt 3}{2}\lim_{p\to0}\frac{a_0(p)}{p}
$$
from the phase shifts. For the HFD-B potential we found
$\ell_{\rm sc}=170{\pm}5$\,{\AA} when only the $l=\lambda=0$ are
taken into account  and $\ell_{\rm sc}=145{\pm}5$\,{\AA} when both
the $l=\lambda=0$ and $l=\lambda=2$ are considered.  We note
here that previous estimate made by Nakaichi-Maeda and Lim
\cite{Nakai} via zero-energy scattering calculations and by
employing a separable approximation for the HFDHE2 potential
gave the value of $\ell_{\rm sc}=195$\,\AA.

It is interesting to compare the  results for $\ell_{\rm sc}$
with the inverse wave numbers $\kappa^{-1}$ for the trimer
excited state energies,
$
   \kappa=\displaystyle\frac{2}{\sqrt{3}}
   \sqrt{\epsilon_d-E_t}
$,
where the trimer excited state and dimer bound state energies
$E_t$ and $\epsilon_d$ are measured in {\AA}$^{-2}$.  For the
HFD-B interaction we find  $\kappa^{-1}\approx 102$\,{\AA} with
$l=\lambda=0$ and $\kappa^{-1}\approx  89$\,{\AA} with
$l=\lambda=0$ and $l=\lambda=2$. These are about 1.7 times
smaller than the above estimates for $\ell_{\rm sc}$. This is
compared  with  the $^4$He two-atomic scattering  results where
the inverse wave number
$\bigl(\kappa^{(2)}\bigr)^{-1}=84.8$\,{\AA} is a good
approximation for the $^4$He--$^4$He scattering length,
$\ell_{\rm sc}^{(2)}=88.6$\,{\AA}.  Such a significant
difference between $\kappa^{-1}$ and $\ell_{\rm sc}$ can be
attributed to the Efimov properites of the trimer system which imply
that the effective range $r_0$ of the interaction between $^4$He 
atom and $^4$He dimer is very large as compared to the $^4$He 
two-atomic problem. Unfortunately, insufficient accuracy of our 
results for the amplitude $a_0(p)$ at $p\approx 0$ which we have 
at the moment does not allow us to extract the values for the 
$r_0$.

It should be noted that the $^4$He$_3$ system is probably one of
the most challenging problems for any three-body scattering
calculation, not only because of the hard-core of the pair forces, 
but also in view of its pre-Efimov nature. The latter
manifests itself in a very slow falling off of the dimer wave
function and then, as a consequence, in very large hyperradius
values for the asymptotical boundary conditions~(\ref{HeBS})
and~(\ref{AsBCPartS}) for the trimer excitede state and
scattering wave functions were fulfilled.  In our
finite-difference calculations we had to increase the cut-off
radius $\rho_{\rm max}$ up $400-600$\,{\AA} while we had to use
grids with up to 600 knots in both hyperradius $\rho$ and
hyperangle $\theta$ variables until the converged results were
obtained.  All this required for storage of the resulting
matrices up to 5\,Gb of a (hard-drive) memory.  Calculation of
each phase shift point was also very expensive in time requiring
in the case of the two equations ($l,\lambda=0,2$) up to ten or
more hours.  We plan to describe more details of our technics in
an extended article which is under preparation.

Our results  clearly demonstrate the reliability of our method
in three-body bound state and scattering calculations in system
where the inter-atomic potential contains a hard-core which
makes such calculations extremely tedious and numerically
unstable. Thus the present formalism paves the way to study
various three-atomic systems, and to calculate important
quantities such as  cross-sections, recombination rates etc.


\bigskip
\acknowledgements
Financial support from the University of South Africa, the Joint
Institute for Nuclear Research, Dubna, and the Russian
Foundation for Basic Research (Projects No.~96-01-01292,
No.~96-01-01716 and No.~96-02-17021) is  gratefully
acknowledged.  The authors are indebted to Dr.~F.~M.~Penkov for
a number of useful remarks and to Prof.~I.~E.~Lagaris for
allowing us to use the computer facilities of the University of
Ioannina, Greece, to perform scattering  calculations.




\begin{table}
\label{tableI}
\caption
{Bound state energy (in K) results for the Helium trimer.}
\begin{tabular}{|c|ccccc|cc|c|}
\hline
Potential & \multicolumn{5}{c|}{Faddeev equations}
&\multicolumn{2}{c|}{Variational}&\multicolumn{1}{c|}{Adiabatic} \\
 & \multicolumn{5}{c|}{}
&\multicolumn{2}{c|}{methods}&\multicolumn{1}{c|}{approach} \\
\cline{2-9}
   & $l$ &  This work & \cite{CGM} & \cite{Gloeckle}
& \cite{Nakai} & \cite{V2} & \cite{V5}
& \multicolumn{1}{c|}{\cite{EsryLinGreene}}\\
\hline \hline
HFDHE2 & 0 & 0.084 &  & 0.082 & 0.092 &  &   & 0.098
\\ \cline{2-6}
 & 0,2 &  0.114 & 0.107 & 0.11 &  & 0.1173 &  &  \\
\hline \hline
 HFD-B & 0 &0.096 & 0.096 & & & & & \\
\cline{2-6}
 & 0,2 & 0.131  & 0.130 & &  & & 0.1193 & \\
\hline
\end{tabular}
\end{table}


\begin{table}
\label{tableII}
\caption
{Excited state energy (in mK) results for the Helium trimer.}
\begin{tabular}{|c|c|c|c|c|c|}
\hline
Potential   & $l$ &  This work &  \cite{Gloeckle} & \cite{Nakai}
					  &\cite{EsryLinGreene}\\
\hline \hline
HFDHE2      &  0  &    1.5     &       1.46    &  1.04 & 1.517  \\
\cline{2-5}
            & 0,2 &    1.7     &       1.6     &   & \\
\hline \hline
 HFD-B      &  0  &    2.5     &               &  &\\
\cline{2-5}
            & 0,2 &    2.8     &               &  &\\
\hline
\end{tabular}
\end{table}



\begin{figure}
\centering
\epsfig{file=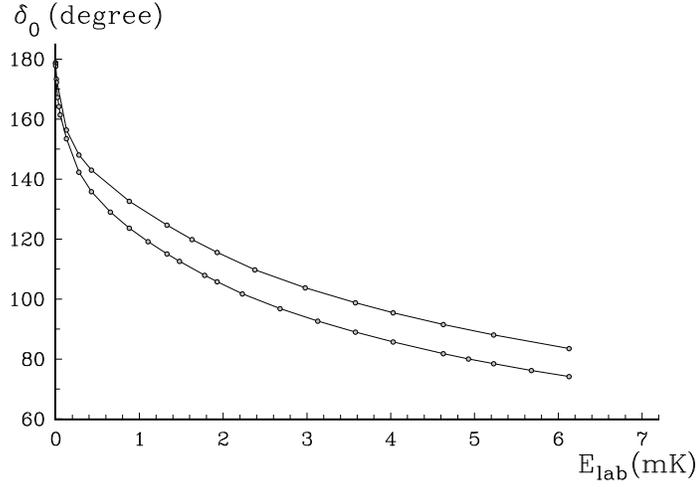,height=8cm}
\caption{S-wave Helium atom -- Helium dimer scattering phase
shifts $\delta_0(E_{\rm lab})$, $E_{\rm lab}=\frac{3}{2}(E+|\epsilon_d|)$,
 for the HFD-B $^4$He--$^4$He
potential. The lower curve corresponds to the case where only
$l=0$ are taken into account while for the upper both $l=0$ and
$l=2$.}
\label{Fig-phases}
\end{figure}


\begin{figure}
\centering
\psfig{file=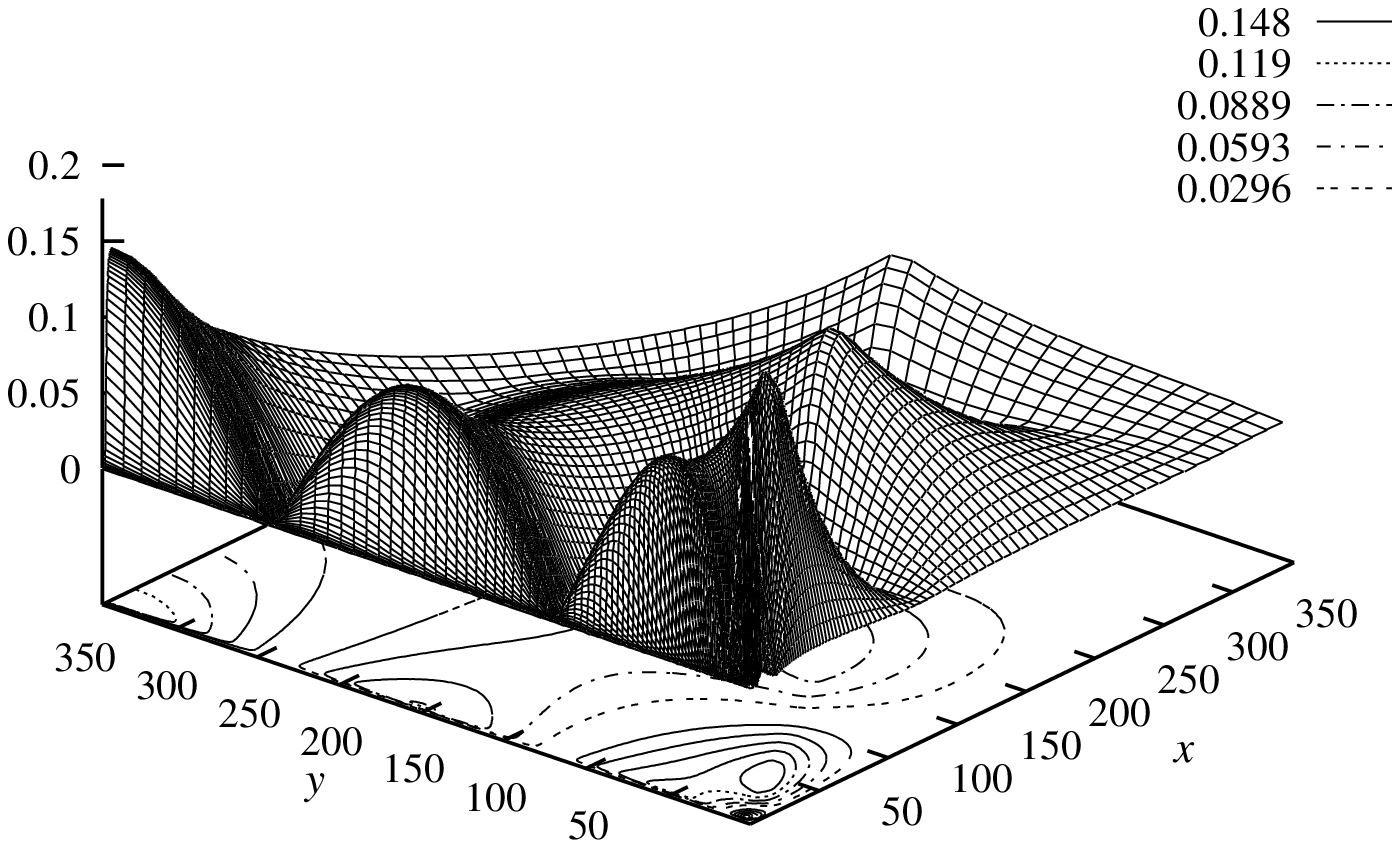,height=7cm}
\caption{Absolute value of the wave function component
$\Psi_0(x,y,p)$ for the HFD-B $^4$He--$^4$He potential
at $E=+1.4$\,mK. Values of $x$ and $y$ are in~\AA.}
\label{wf1a}
\end{figure}


\begin{figure}
\centering
\psfig{file=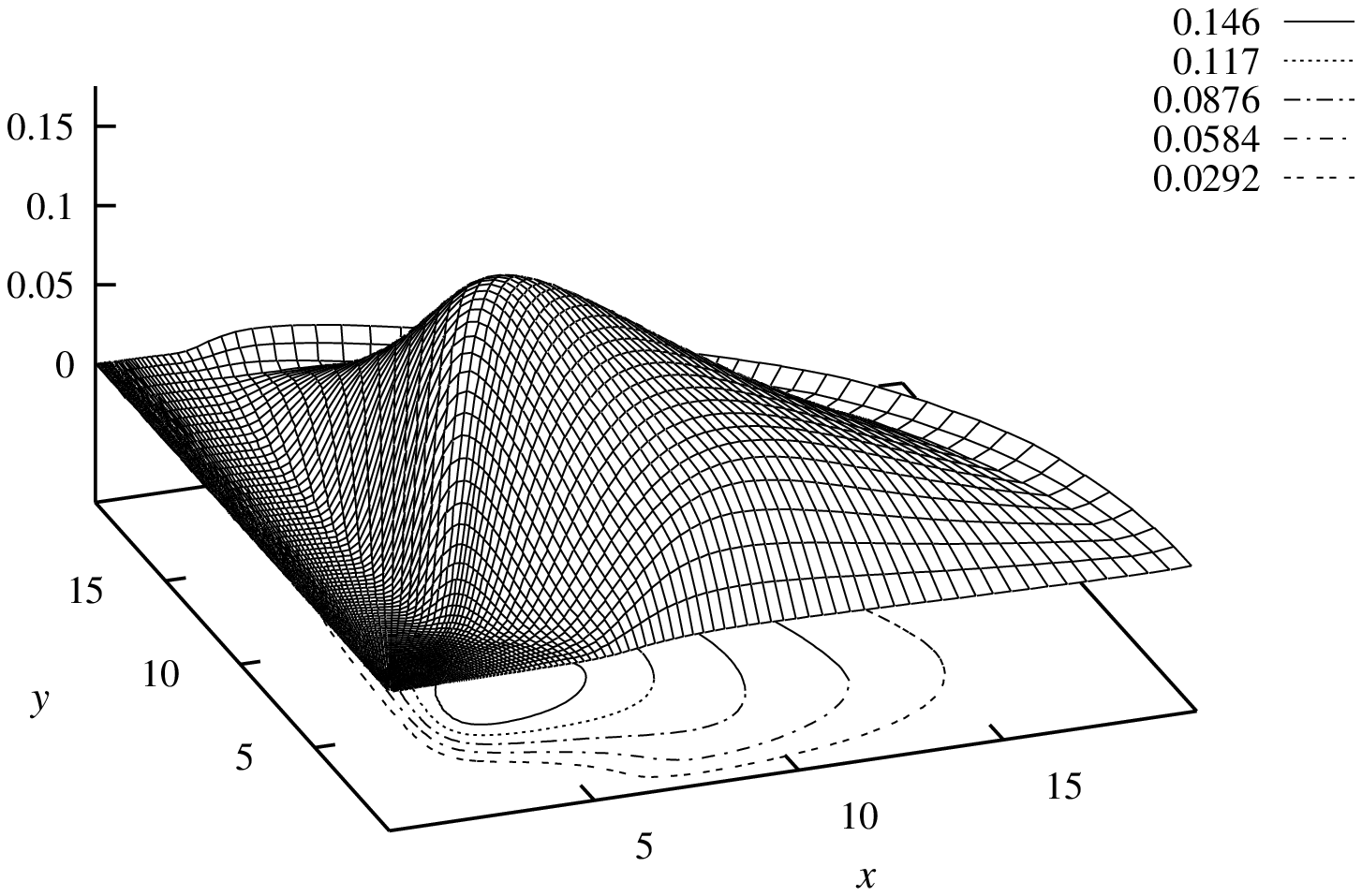,height=5.5cm}
\caption{Detail of the $|\Psi_0(x,y,p)|$ surface
shown in Fig.~2.}
\label{wf1dp}
\end{figure}


\begin{figure}
\centering
\psfig{file=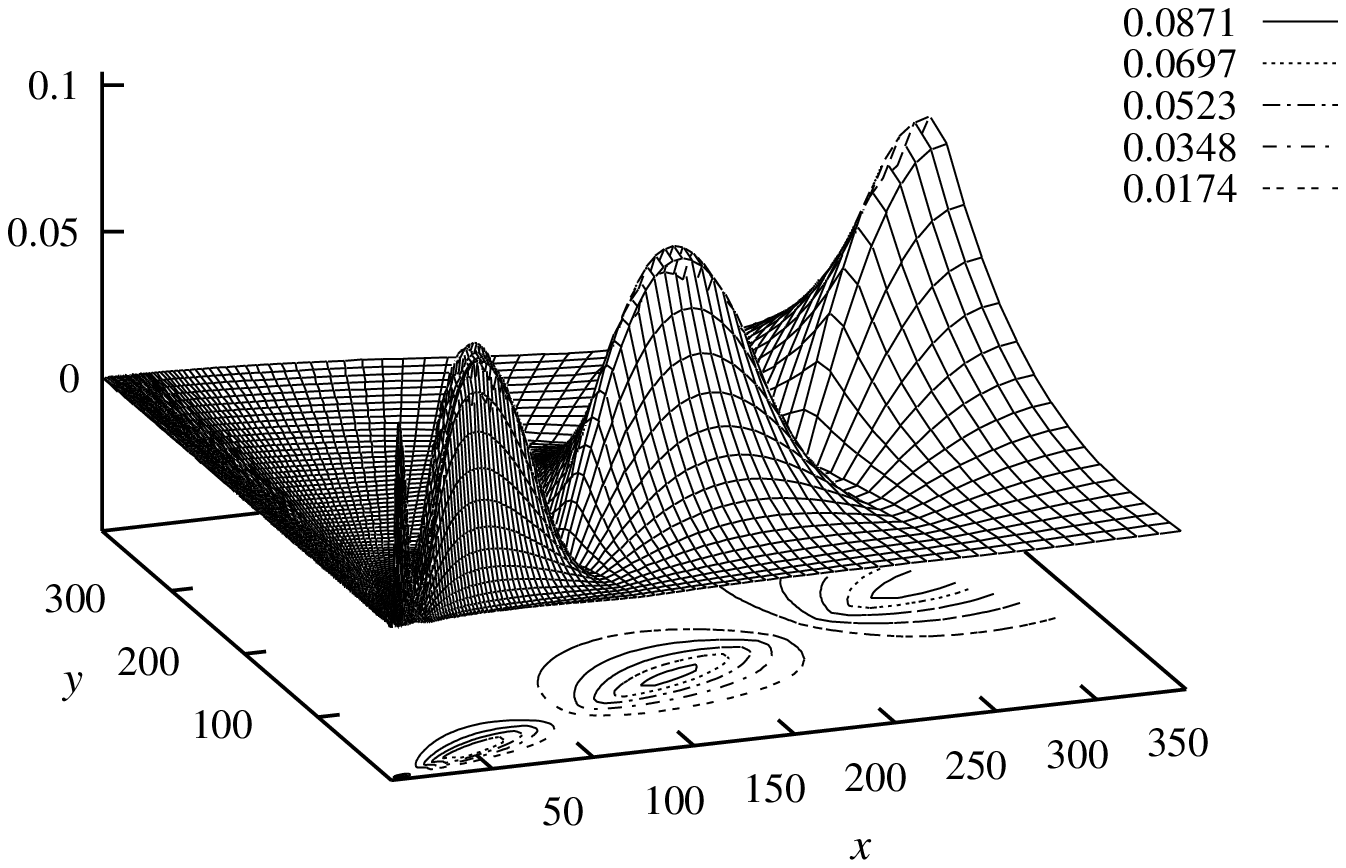,height=7cm}
\caption{Absolute value of the wave function component
$\Psi_2(x,y,p)$ for the HFD-B $^4$He--$^4$He potential
at $E=+1.4$\,mK. Values of $x$ and $y$ are in~\AA.}
\label{wf2a}
\end{figure}


\begin{figure}
\centering
\psfig{file=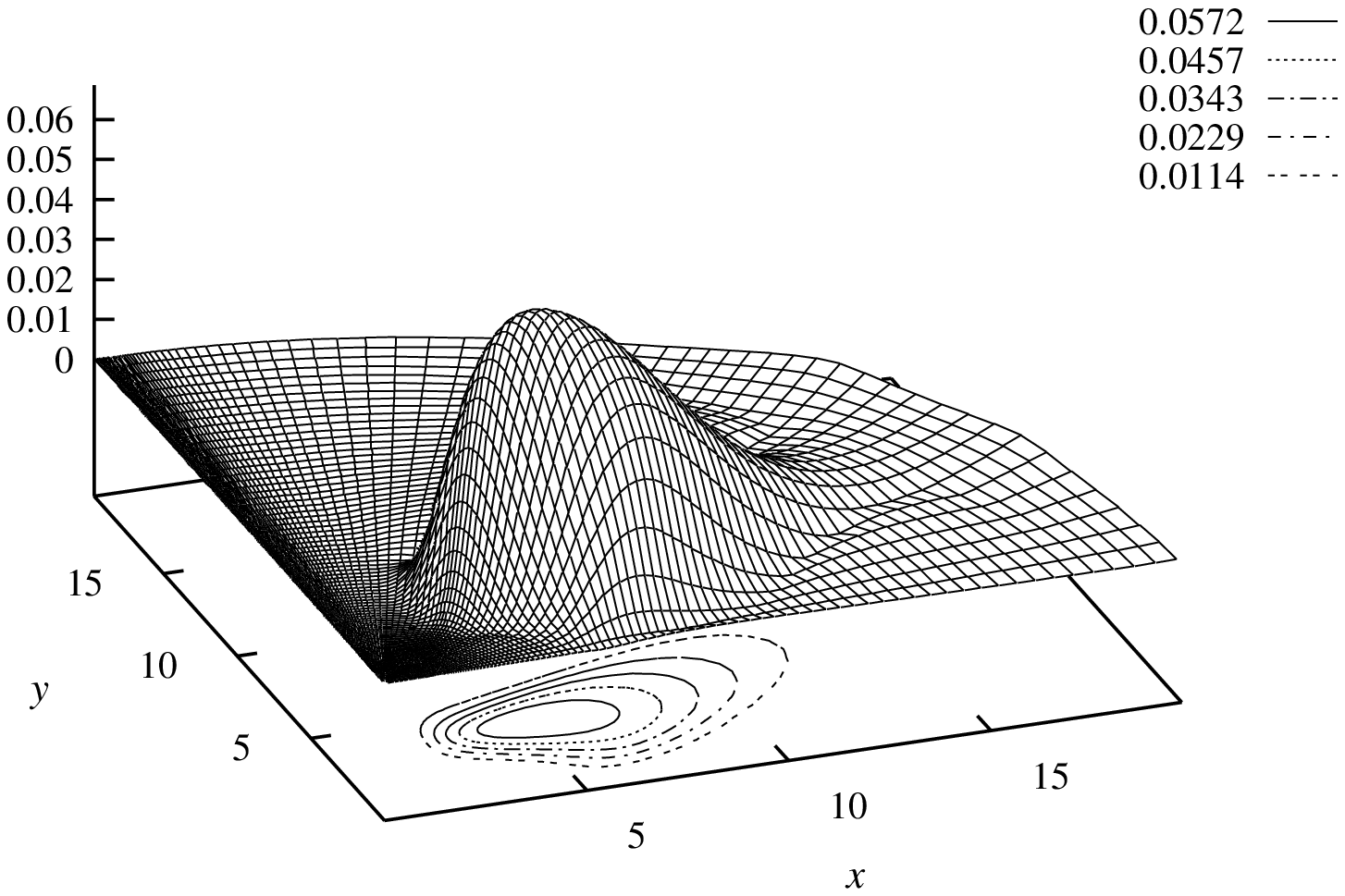,height=5.5cm}
\caption{Detail of the $|\Psi_2(x,y,p)|$ surface
shown in Fig.~4.}
\label{wf2dp}
\end{figure}

\end{document}